\documentclass[aps,pra,preprint,superscriptaddress,longbibliography]{revtex4-1}

\usepackage{amsmath}
\usepackage{amsthm}
\usepackage{amsfonts}
\usepackage{amssymb}
\usepackage{xcolor,graphicx}
\usepackage{tikz}
\usepackage{color}
\usepackage[pdftex]{hyperref} 
\usepackage{longtable}
\usepackage{array}
\usepackage{dsfont}

\begin{document}
	
	\title{Noise-assisted routing in mode-coupling devices}
	
	\author{P. Bravo-Cassab}
	\affiliation{Benem\'erita Universidad Aut\'onoma de Puebla, Facultad de Ciencias Fisico Matemáticas, Av. San Claudio, Cd Universitaria, Jardines de San Manuel, Puebla, Pue., Mexico, 72572}
	
	\author{B. Jaramillo-\'Avila}
	\email[e-mail: ]{jaramillo@inaoep.mx}
	\affiliation{CONACYT - Instituto Nacional de Astrof\'isica, \'Optica y Electr\'onica, Calle Luis Enrique Erro No. 1. Sta. Ma. Tonantzintla, Pue., Mexico, 72840.}

	\author{B. M. Rodr\'iguez-Lara}
	\email[e-mail: ]{bmlara@tec.mx}
	\affiliation{Tecnologico de Monterrey, Escuela de Ingenier\'ia y Ciencias, Ave. Eugenio Garza Sada 2501, Monterrey, N.L., Mexico, 64849.}	
	
	\date{\today}
	
	\begin{abstract}
	The dynamical control of energy transfer between interacting systems is fundamental in diverse applications related to physical, electronic and chemical processes.
	Recent developments show that noise may enhance or suppress power transfer in systems described by Coupled-Mode Theory.
	We show a semi-analytic approach to utilize dynamical noise to produce routing in coupled mode devices. 
	We present results for an optical mode-coupling device with induced noise in the refractive index of the material.
	However, our approach is valid for networks of coupled oscillators of any type as long as their dynamics are described by Coupled-Mode Theory. 
	\end{abstract}
	
	
	\maketitle

\section{Introduction}
Noise is a common feature in physical systems, where it is often disruptive. 
However, an adequate amount of noise may induce novel and useful effects; for example, the effect of noise on energy transport is a particularly interesting. It may be observed in optical, quantum and electronic systems, as the frameworks describing these systems are often connected by analogies. 

Noise-induced effects may appear as a consequence of either constant or dynamical perturbations. 
The former remain constant throughout the evolution or propagation and is behind Anderson localization \cite{Anderson1958,Anderson1985,Lagendijk2009}, where random changes in the parameters of the system produce localization of light beams \cite{John1984,John1987,DeRaedt1989,Yannopapas2003,Schwartz2007,Martin2011,Mafi2015}, electron quantum wavefunctions \cite{Lee1981,Gornyi2005,Ying2016}, and even atoms in Bose-Einstein condensates \cite{Roati2008}.
On the other hand, dynamical noise randomly changes as the system evolves and may enhance or suppress the transfer of quantum and classical excitations; for example, noise-assisted transport in photosynthetic complexes, where dynamical noise helps high-efficiency excitation transfer \cite{Plenio2008,Caruso2009}. 
In optical systems, it is possible to realize noise-assisted transport \cite{Viciani2015,Viciani2016,Biggerstaff2016,Caruso2016,Intravaia2019} or suppress crosstalk \cite{Takenaga2011,Matsuo2011a,Matsuo2011b,Fini2012,Fini2010,Li2015,Chen1999,Jaramillo2019a}.

Here, we focus on noise-assisted routing of signals propagating through systems described by the Coupled-Mode Theory framework. 
These include but are not limited to optical waveguides \cite{Snyder1972,McIntyre1973,Huang1994}, 
Terahertz resonators \cite{Preu2008}, microwave cavities \cite{Haus1991,Elnaggar2014,Elnaggar2015}, RF antennas \cite{Kim2007}, RLC circuits \cite{Agarwal2006}, and microring resonators \cite{Liu2005}.
Furthermore, Coupled-Mode Theory helps building an analogy between classical systems, as those described before, and single-excitation discrete quantum systems \cite{Longhi2009}.
In particular, we use light propagation through an array of coupled waveguiding cores as physical platform to show signal routing induced by differentiated dynamical noise in the refractive index of the cores along the propagation direction.
Laser writing of waveguides is a technique that allows the fabrication of devices with sufficient control over the refractive index of the cores \cite{Davis1996,Blomer2006,Eaton2011}. 
A drawback of this platform is the smallness of attainable reconfigurable noise. 
Electronics may be a more suitable platform where reconfigurable dynamical noise is easier to produce and control at any given scale using RLC coupled circuits \cite{Leon2015,Quiroz2016,Leon2018}. 
In the next section, we introduce our system. For the sake of simplicity, we induce an underlying symmetry to ease the design and modelling process.
This allows us to produce a semi-analytical description of propagation under noisy conditions. 
Next, we present the results of our numerical experiments and optimize the amplitude of the noise to induce signal routing. 
Finally, we close with our conclusions.

\section{Model}

\begin{figure*}
	\includegraphics{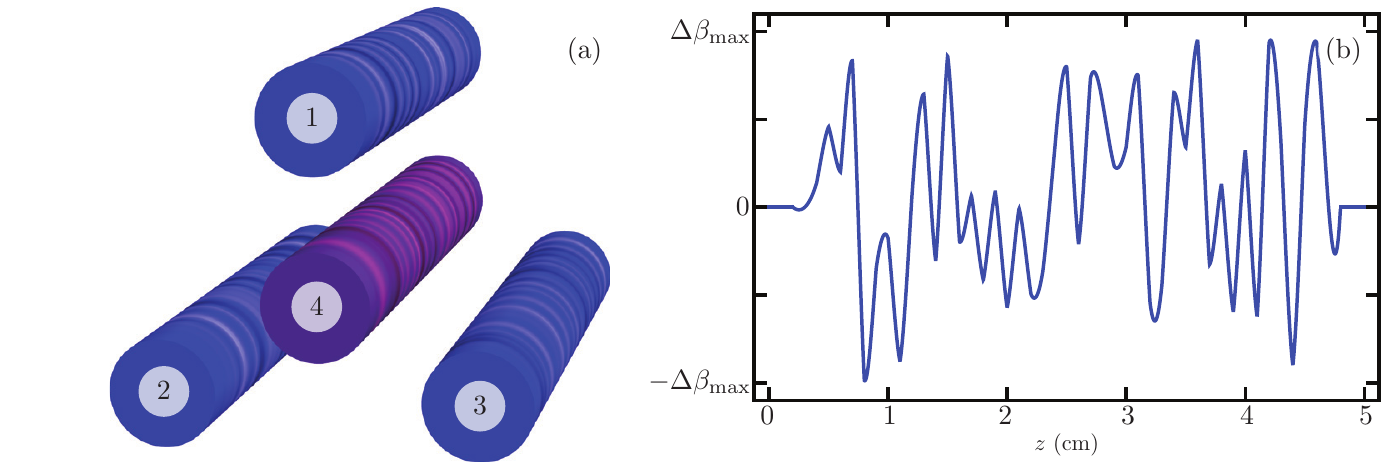}
	\caption{(a) Sketch of a four-core realization of our proposal. 
		Three nearly identical cores are homogeneously distributed around a central core. The refractive index of each core is modulated to include a random noise perturbation that independently modifies (b) the effective propagation constant of each core as a function of the propagation length $z$.}\label{fig:Fig1}
\end{figure*}

For the sake of providing a tractable methodology, we focus on an waveguide array with a well defined symmetry that allows calculating the normal modes of the unperturbed system. 
We consider a necklace of $N$ identical equidistant cores around a central core \cite{Jaramillo2019b}. 
In particular, we present results for a four-core realization, Fig. \ref{fig:Fig1}(a), but the method works for larger systems.
The refractive index of each core is perturbed independently by a random fluctuation that is a continuous and differentiable function of the propagation variable $z$ that, in consequence, changes the effective propagation constant of the localized mode at each core; for example, Fig. \ref{fig:Fig1}(b).
The dynamics of the system is governed by the coupled mode equation, \begin{align}
	- i \frac{\text{d}}{\text{d}z} \vec{\mathcal{E}}(z) = \mathbf{M}(z) \cdot \vec{\mathcal{E}}(z),
\end{align}
where the scalar complex field amplitudes of the localized modes are stored in the $N$-dimensional vector $\vec{\mathcal{E}}(z)$ and the information of the effective optical system in the coupled mode matrix $\mathbf{M}(z)$. We split the latter,
\begin{align}
	\mathbf{M}(z) = \mathbf{M}_{0} + \delta \mathbf{M}(z),
\end{align}
into a dominant unperturbed constant part $\mathbf{M}_{0}$ and a smaller $z$-dependent random fluctuation part $\delta \mathbf{M}(z)$.
The symmetry of this array is the $N$-element cyclic group $\mathbb{Z}_{N}$ \cite{Jaramillo2019b}.
In our particular four-core example, the constant part of the coupled mode matrix reflects this symmetry,
\begin{align}
	\mathbf{M}_{0} &= 
	\left(
	\begin{array}{cccc}
		\beta	& g		& g		& g_{c}		\\
		g		& \beta	& g		& g_{c}		\\
		g		& g		& \beta	& g_{c}		\\
		g_{c}	& g_{c}	& g_{c}	& \beta_{c}
	\end{array}
	\right),
\end{align}
where the effective propagation constant of the localized modes in the external (central) cores is $\beta$ ($\beta_{c}$), the effective constant coupling strength between external cores is $g$ and between external and central cores is $g_{c}$.
Additionally, we consider $z$-dependent fluctuations that modify the core refractive indices.
A change in the refractive indices of the cores induces a change in both the effective propagation constant and the coupling strengths. 
We consider fluctuations small enough to slightly change the effective propagation constant at each core while producing a negligible change in the coupling strengths \cite{Jaramillo2019a}, such that the fluctuation matrix is diagonal,
\begin{align}
	\delta \mathbf{M}(z) &= \text{diag}\left[ \delta\beta_{1}(z), \delta\beta_{2}(z), \delta\beta_{3}(z), 0 \right],
\end{align}
and we restrict the noise to external cores for the sake of simplicity. 

The unperturbed case, $\mathbf{M} = \mathbf{M}_{0}$ with $\delta \mathbf{M}(z) = 0$, accepts analytic normal modes \cite{Jaramillo2019b,Jaramillo2019a} whose propagation constants have the form,
\begin{align}
\begin{aligned}
	\lambda_{1} &= \lambda_{2} = \beta - g, \qquad \\
	\lambda_{3} &= \frac{1}{2} \left[ \lambda_{0} + \beta_{c} - \sqrt{ \left( \lambda_{0} - \beta_{c} \right)^{2} + 12 g_{c}^{2} } \right], \qquad \\
	\lambda_{4} &= \frac{1}{2} \left[ \lambda_{0} + \beta_{c} + \sqrt{ \left( \lambda_{0} - \beta_{c} \right)^{2} + 12 g_{c}^{2} } \right],
\end{aligned}
\end{align}
where we use the auxiliary quantity $\lambda_{0} = \beta + 2 g$ for the sake of space.
The corresponding normal modes are written in terms of an extension of the three-dimensional Fourier matrix,
\begin{align}
\left[ \mathbf{F}_{e} \right]_{i,j} = 
\left\{\begin{array}{ll}
	e^{2\pi i (i-1)(j-1)/3}/\sqrt{3}	& \text{ for } 1 \leq i \leq 3 \text{ and } 1 \leq j \leq 3		\\
	\delta_{i,j}						& \text{ for } i=4 \text{ or } j=4
\end{array}
\right.,
\end{align}
and have the form
\begin{alignat}{3}
\begin{aligned}
	\hat{v}_{1} &= \mathbf{F}_{e} \cdot \hat{e}_{2},& \qquad
	\hat{v}_{2} &= \mathbf{F}_{e} \cdot \hat{e}_{3},& \\
	\hat{v}_{3} &= - \sin \theta ~\mathbf{F}_{e} \cdot \hat{e}_{1} + \cos \theta ~\hat{e}_{4},& \qquad
	\hat{v}_{4} &= \cos \theta ~\mathbf{F}_{e} \cdot \hat{e}_{1} + \sin \theta ~\hat{e}_{4},&
\end{aligned}
\end{alignat}
where the unitary vectors $\hat{e}_{k}$ are the standard canonical basis, with value of 1 in the $k$-th entry and 0 everywhere else, 
and the mixing angle fulfils, 
\begin{align}
	\tan \theta = \frac{2\sqrt{3} g_{c}}{\lambda_{0}-\beta_{c}+\sqrt{\left(\lambda_{0}-\beta_{c}\right)^{2}+12g_{c}^{2}}}.
\end{align}
The system has a two-dimensional removable degeneracy in the first two effective propagation constants, $\lambda_{1}$ and $\lambda_{2}$. 
This makes the standard approach for constant, or $z$-independent, first-order perturbations inadequate.
The degeneracy produces a divergence in denominators with the form $\lambda_{1}-\lambda_{2}$. 
To prevent this, we find two new orthonormal eigenvectors that lie within the degenerate subspace, i.e., the space spanned by $\hat{v}_{1}$ and $v_{2}$, and diagonalize the perturbation \cite{BookShankarPrinciples}. These new vectors also form an orthonormal set of eigenvectors for the unperturbed coupled mode matrix, $\mathbf{M}_{0}$, and are as valid as the original ones. 
However, as these new vectors diagonalize the perturbation within the degenerate subspace, $\hat{w}_{1}^{\dagger} \cdot \delta\mathbf{M} \cdot \hat{w}_{2} = 0$, they are safe to use in standard first-order perturbation theory.
These new eigenvectors have the form,
\begin{alignat}{3}
\begin{aligned}
	\hat{w}_{1} &= \hat{v}_{1}/\sqrt{2} + e^{-i \phi}~ \hat{v}_{2}/\sqrt{2},& \qquad
	\hat{w}_{2} &= -e^{i \phi}~ \hat{v}_{2}/\sqrt{2} + \hat{v}_{2}/\sqrt{2},& \\
	\hat{w}_{3} &= \hat{v}_{3},& \qquad
	\hat{w}_{4} &= \hat{v}_{4},&
\end{aligned}
\end{alignat}
where we define a phase in terms of the perturbations,
\begin{align}
	\phi &= \text{arg}\left( 
		\frac{
			\delta\beta_{1} + (-1)^{2/3} \delta\beta_{2} - (-1)^{1/3} \delta\beta_{3}
		}{
			\delta\beta_{1}^{2}+\delta\beta_{2}^{2}+\delta\beta_{3}^{2}-\delta\beta_{1}\delta\beta_{2}-\delta\beta_{1}\delta\beta_{3}-\delta\beta_{2}\delta\beta_{3}
		}
	\right).
\end{align}
Now, assuming constant perturbations, the first-order perturbed vectors have the form
\begin{align}
\vec{a}_{j} = \hat{w}_{j} + \sum_{k=1, k \neq j}^{4} \frac{\hat{w}_{k}^{\dagger} \cdot \delta\mathbf{M} \cdot \hat{w}_{j}}{\lambda_{j}-\lambda_{k}} \hat{w}_{k},
\end{align}
where terms with denominators $\lambda_{1}-\lambda_{2}$ have a vanishing numerator. 
These perturbed eigenvectors are not normalized, so they need to be normalized again. 

We are interested in dynamical, $z$-dependent, noise in the waveguides. On the other hand, the above expressions for perturbed vectors apply to constant, $z$-independent, perturbations. However, these expressions become useful to describe propagation over long distances, where light passes through many different and independent fluctuations. This acts as an averaging process and we are able to use statistical properties of the noise in the expressions for constant perturbations \cite{Jaramillo2019a}. In particular, we replace powers of the perturbations $\delta \beta_{i}^{p}$ for the corresponding momenta of the noise distributions $\langle \delta \beta_{i}^{p} \rangle$.

 We optimize noise routing along the $k$-th core by calculating the inverse participation ratio (IPR) \cite{Thouless1974,Wegner1980} of the corresponding vector $\hat{e}_{k}$.
 We use the basis of normalized and perturbed vectors $\left\{\hat{a}_{j}\right\}_{j=1,\ldots,4}$, such that
\begin{align}
\text{IPR}\left( \hat{e}_{k}, \left\{\hat{a}_{j}\right\}_{j=1,\ldots,4} \right) = \sum_{j=1}^{4} \big\vert \hat{e}_{k}^{\dagger} \cdot \hat{a}_{j} \big\vert^{4}.
\end{align}
If this IPR approaches its maximum value of one, light mostly propagates through the $k$-th core. 
Therefore, the optimization consists on bringing the IPR as close as possible to one.
An exact normalization of the perturbed vectors $\vec{a}_{j}$ produces complicated expressions. However, they form an orthonormal basis up to first order and, in consequence, we use a series expansion for the terms that normalize the vectors $1/|| \vec{a}_{j}||$.
Despite their complicated form involving hundreds of terms, the inverse participation ratios for the external cores are symmetric; that is, the expressions are identical under cyclic substitutions and reflections in the indices of the three external cores,
\begin{align}\label{eq:indexsymmetry}
\text{IPR}\left( \hat{e}_{1} \right)
&= \left.\text{IPR}\left( \hat{e}_{3} \right)\right|_{1 \to 2,~ 2 \to 3,~ 3 \to 1}
= \left.\text{IPR}\left( \hat{e}_{2} \right)\right|_{1 \to 3,~ 2 \to 1,~ 3 \to 2}
= \left.\text{IPR}\left( \hat{e}_{1} \right)\right|_{2 \leftrightarrow 3}
= \left.\text{IPR}\left( \hat{e}_{2} \right)\right|_{1 \leftrightarrow 3}
\nonumber \\
&=\left.\text{IPR}\left( \hat{e}_{3} \right)\right|_{1 \leftrightarrow 2}.
\end{align}
The first two identities correspond to clockwise cyclic rotations and the last three correspond to reflections along the symmetry axes of the array.
Additionally, these expressions are significantly simplified once we substitute for numerical values of the system parameters. 

\section{Results}
We want to stress that our approach is useful for any physical platform described by Coupled-Mode Theory.
Nonetheless, we focus on light propagation through an array of coupled waveguiding cores due to our experience. 
We consider four identical cores as the zero noise system. 
The cores have a radii of $4.5~\mu\textrm{m}$ and refractive index $1.447\,9$. We assume a cladding with refractive index of $1.444\,0$ and use light in the telecomm C-band with wavelength $\lambda = 1550~\mathrm{nm}$. 
These conditions provide an effective propagation constant of $5.859\,8 \times 10^{6}~\textrm{rad}/\textrm{m}$ for the single guided field mode at each core. 
The three external cores are placed $15~\mu\textrm{m}$ away from the central core and $25.980\,8~\mu\textrm{m}$ from each other. 
These are center-to-center distances producing cooupling strengths of $256.635\,5~\textrm{rad}/\textrm{m}$ and $8.252\,0~\textrm{rad}/\textrm{m}$ between external and central core and pairs of external cores, in that order.
We choose this configuration to explore a regime that produces strong crosstalk between guided modes in the external cores compared with that to the central core.
This will help us determine that routing arises from noise and not from the symmetry of the system.
However, this selection makes the optical uses of our approach impractical as the propagation distances required to see an effect become of the order of a meter.
Coupled electronic oscillators may be a better platform to explore this effect in a laboratory \cite{Leon2015}. 

We work with a first-order perturbation expansion and substitute powers of the constant noise by their momenta.
This and the optical parameters allow us to obtain a simpler expression for the inverse participation ratios.
Furthermore, if we consider that positive and negative noise is equally probable, all the odd-powered momenta vanish. 
In consequence, the inverse participation ratios have the form
\begin{align}
\begin{aligned}\label{eq:iprsnum}
\text{IPR}\left( \hat{e}_{1},\left\{ \hat{a}_{j} \right\}_{j=1,\ldots,4} \right) 
\to \;&
\big[ 
	0.500\,0~ m_{1} 
	+ 0.333\,4~ (m_{2}+m_{3}) 
	+ 0.037\,1~ m_{1}~ (m_{2}+m_{3}) 
\\
&	- 0.111\,3~ m_{2}~ m_{3}
\big]/(m_{1}+m_{2}+m_{3})
\\
\text{IPR}\left( \hat{e}_{4},\left\{ \hat{a}_{j} \right\}_{j=1,\ldots,4} \right) 
\to \;&
0.500\,2 - 0.032\,6~ (m_{1}+m_{2}+m_{3}),
\end{aligned}
\end{align}
where we define an squared noise to coupling constant ratio $m_{i} = \langle \delta \beta_{i}^{2} \rangle/g_{c}^{2}$. 
It is possible to obtain the inverse participation ratios for the other cores, $k=2,3$, from the symmetry properties in Eq. (\ref{eq:indexsymmetry}). 
In general, the inverse participation ratio must be bounded between $1/N$, where $N$ is the number of elements in a basis for the space, and one. Therefore, the expressions above, remain valid for small values of the perturbations. 

We use noise produced by refractive index variations of up to $50\%$ of the contrast between cladding and cores. This produces changes of up to $\delta \beta_{\text{max}} / g_{c} = 18.985\,0$. 
We use noise with spatial frequency of $10^{3}~\textrm{m}^{-1}$, obtained from the numerical optimization in Ref. \cite{Jaramillo2019a}. 
As a first example with these parameters, we perform a cohort of independent and random experiments where we add noise to some of the external cores. 
In particular, Fig. \ref{fig:Fig1} displays the averages and standard deviations of irradiance for a sample of 1000 independent experiments with noise added to the third waveguide. 
The unperturbed waveguides show an average irradiance of about $16.65\%$ of the total each and the central and target waveguides of about $33.35\%$ and $33.35\%$.
It may be possible to engineer a system where noise produces particular routing ratios.

\begin{figure}
	\includegraphics{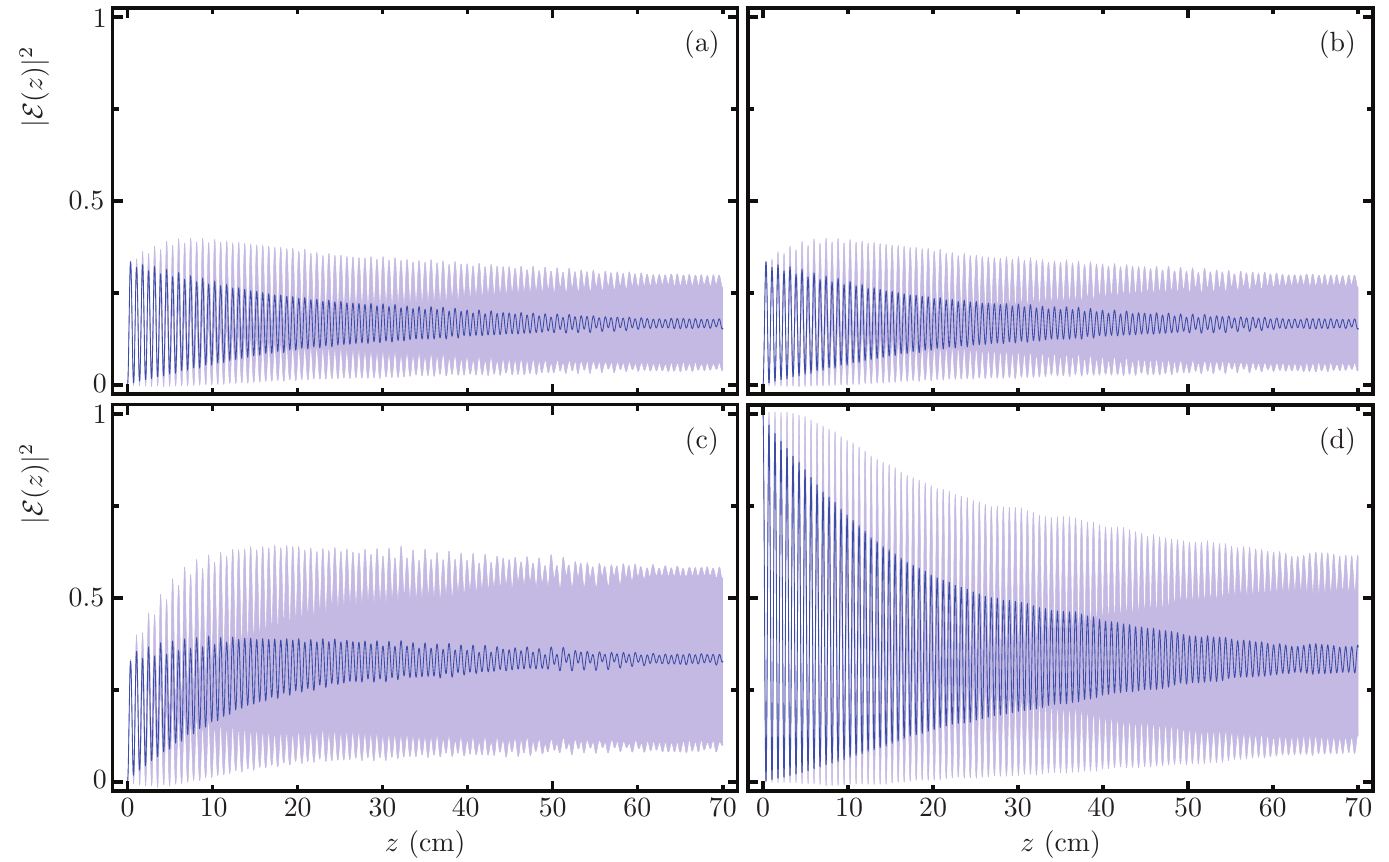}
	\caption{Irradiance as a function of the propagation distance in cores 1 to 4, corresponding to panels (a) to (d) in that order. We add noise to just the third waveguide. The solid line shows the average irrradiance, at each point in $z$, from a sample of 1000 independent and random repetitions with noise in just the third core. The light blue region displays one standard deviation, again for each value in $z$, around the average.}\label{fig:Fig2}
\end{figure}

\begin{figure}
	\includegraphics{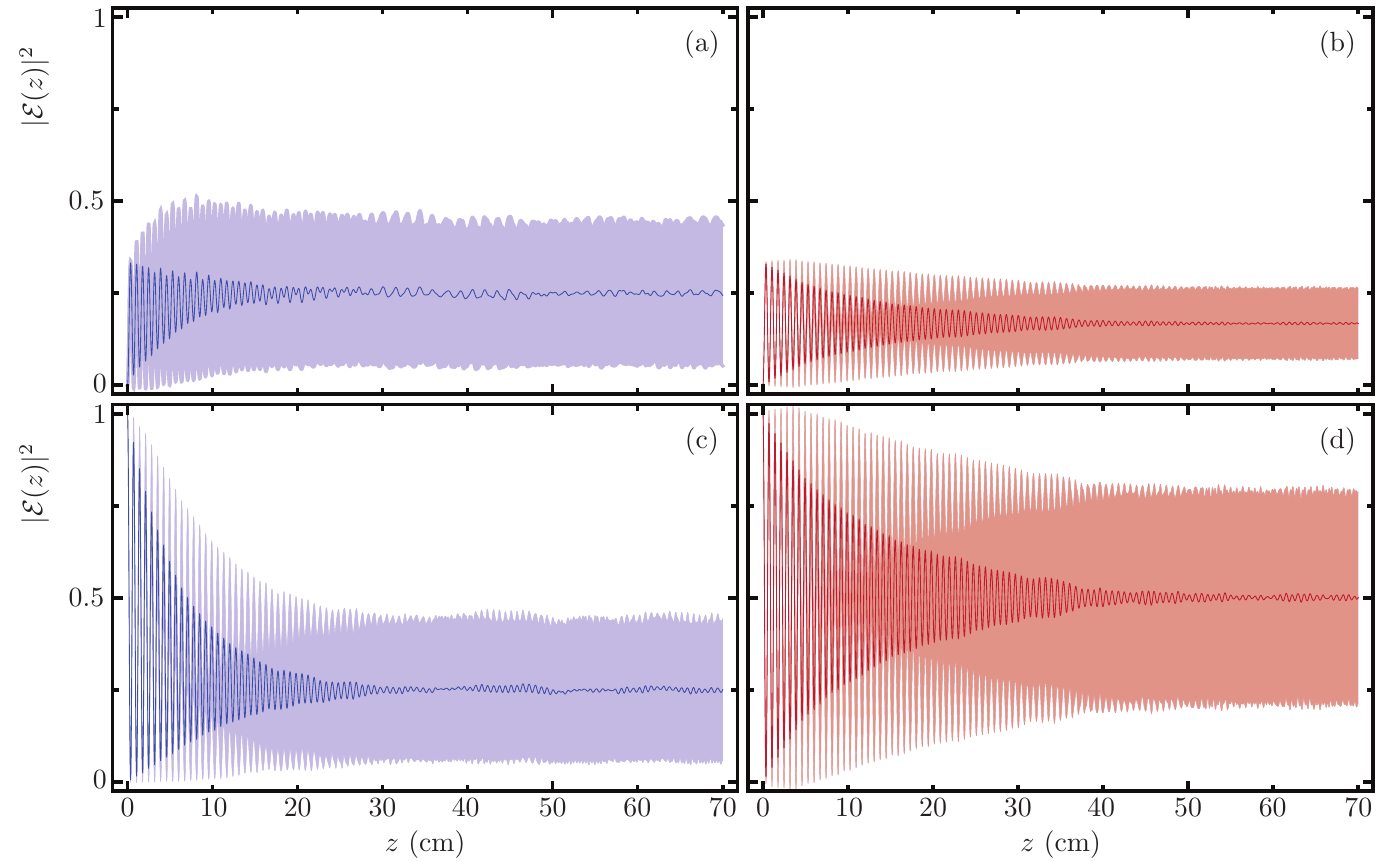}
	\caption{Comparison between adding noise to each external core, left column, v.s. adding noise to the central core, right column. Each panel displays the irradiance as a function of the propagation distance. Panels (a) and (b) display irradiance in core 1, which is very similar to cores 2 and 3, and panels (c) and (d) display irradiance in the external core. As in Figure \ref{fig:Fig2}, the solid line shows the average irrradiance, at each point in $z$, from the sample of independent experiments. The light region displays one standard deviation around the average.}\label{fig:Fig3}
\end{figure}

As an additional numerical experiment, we compare the effect of adding noise to each of the external cores with the effect of adding noise to just the central one. 
As in the previous experiment we use noise with a maximum amplitude of $\delta \beta_{\text{max}} / g_{c} = 18.985\,0$ and spatial frequency of $10^{3}~\textrm{m}^{-1}$. 
We numerically simulate a cohort of 1000 independent experiments and find that adding noise to the external cores increases the fraction of irradiance that propagates through these. 
The three external cores and the central core carry around $25.00\%$ of the irradiance each. 
In contrast, adding noise to just the central core increases the fraction of irradiance that propagates through it. 
In this case the three external cores carry $16.65\%$ of the irradiance each, while the central core carries $50.05\%$. 
This is consistent with the results for the inverse participation ratio in Eq. (\ref{eq:iprsnum}), where increasing noise in a core tends to increase its participation in the perturbed states. 
These results are displayed in Fig.~\ref{fig:Fig3}.

\begin{figure}
	\includegraphics{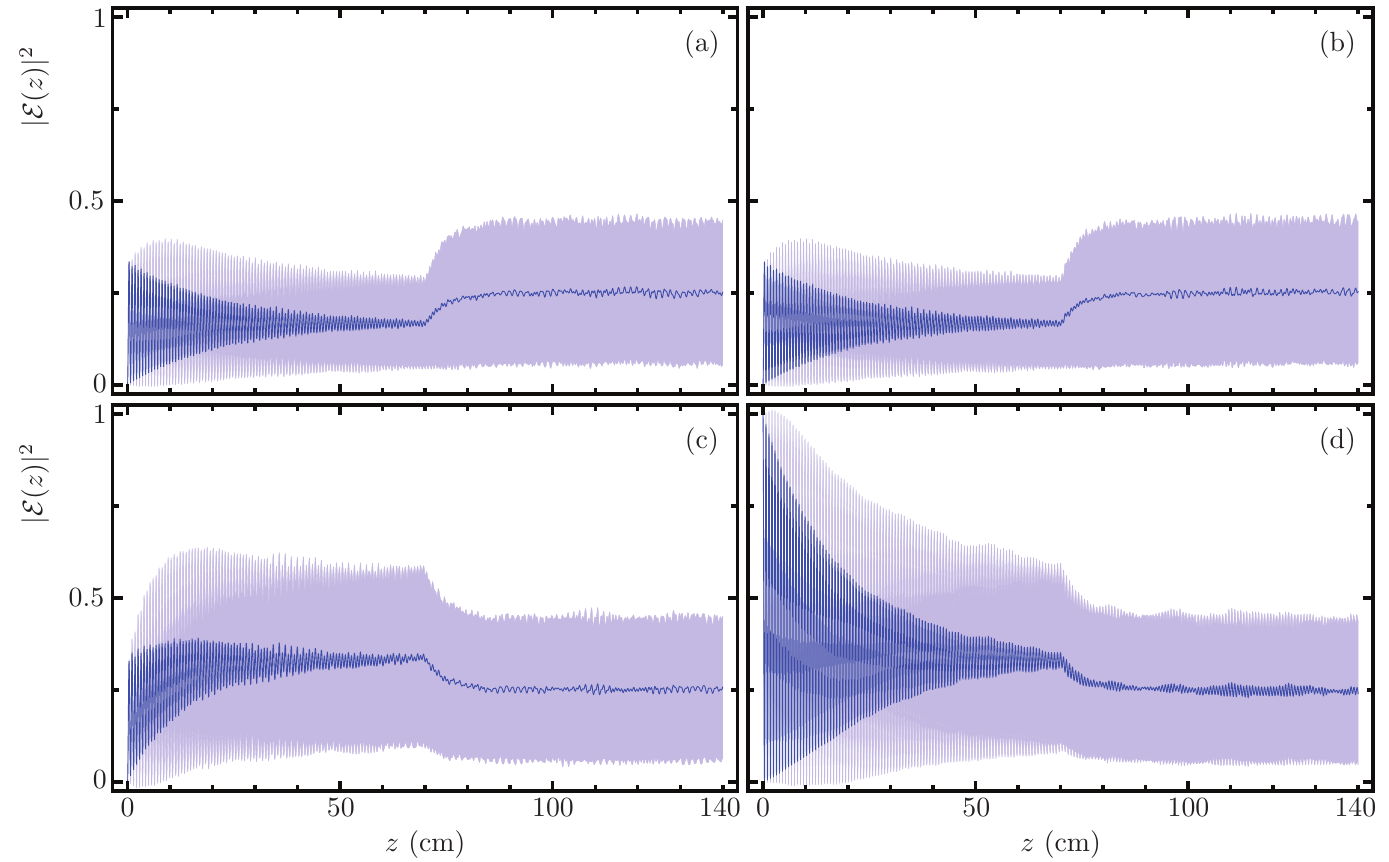}
	\caption{Field irradiance as a function of the propagation distance at cores (a) one, (b) two, (c) three, and (d) four for the numerical experiment with induced noise in the third core that is turnedd off and shifted to cores one and two. Noise is shifted at the $70$ cm mark. Again, solid lines give the average of the cohort and clear regions display one standard deviation above or below the average.}\label{fig:Fig4}
\end{figure}

As a final numerical example, we study the case where noise is added to just one of the external cores and halfway through the propagation it is turned off in the initial core and turned on in the other two external cores. As in the previous experiments, we simulate a cohort of 1000 independent samples and establish noise with a maximum amplitude of $\delta \beta_{\text{max}} / g_{c} = 18.985\,0$ and spatial frequency of $10^{3}~\textrm{m}^{-1}$. 
The results of this numerical experiment are displayed in Fig.~\ref{fig:Fig4}, which show a strong rerouting of irradiance. 
For propagation distances slightly before the midpoint, irradiance is distributed as $16.65\%$ in core 1, $16.65\%$ in core 2, $33.35\%$ in core 3 and $33.35\%$ in core 4, just like in Fig.~\ref{fig:Fig2}. However, after noise is switched to the first two cores, irradiance is distributed as $25.21\%$ in core 1, $25.21\%$ in core 2, $24.99\%$ in core 3 and $24.60\%$ in core 4.

\section{Conclusions}
This work proposes a semi-analytic approach to use diagonal noise for signal routing in systems described by Coupled-Mode Theory. 
We focus on the effect of independent dynamic noise added to individual cores and show that controlling the noise amplitude allows routing the average irradiance to the desired core. 

We perform numerical experiments to demonstrate signal routing in optical waveguides using feasible noise amplitudes under static femtosecond laser writing.
Of course, the lack of reconfigurability is a drawback of the platform. Another is the long propagation distances required to produce effective signal routing. 
However, our approach is valid for any system described by Coupled-Mode Theory. 
We want to stress that networks of electronic oscillators may prove the best candidate to realize reconfigurable, dynamic routing protocols in the laboratory.

\begin{acknowledgments}
P.B.-C. acknowledges financial support from Tec de Monterrey, BUAP and Programa DELFIN 2020. B.J.-A. acknowledges financial support from CONACYT C\'atedra Grupal \#551.
\end{acknowledgments}


\begin{thebibliography}{50}%
	\makeatletter
	\providecommand \@ifxundefined [1]{%
		\@ifx{#1\undefined}
	}%
	\providecommand \@ifnum [1]{%
		\ifnum #1\expandafter \@firstoftwo
		\else \expandafter \@secondoftwo
		\fi
	}%
	\providecommand \@ifx [1]{%
		\ifx #1\expandafter \@firstoftwo
		\else \expandafter \@secondoftwo
		\fi
	}%
	\providecommand \natexlab [1]{#1}%
	\providecommand \enquote  [1]{``#1''}%
	\providecommand \bibnamefont  [1]{#1}%
	\providecommand \bibfnamefont [1]{#1}%
	\providecommand \citenamefont [1]{#1}%
	\providecommand \href@noop [0]{\@secondoftwo}%
	\providecommand \href [0]{\begingroup \@sanitize@url \@href}%
	\providecommand \@href[1]{\@@startlink{#1}\@@href}%
	\providecommand \@@href[1]{\endgroup#1\@@endlink}%
	\providecommand \@sanitize@url [0]{\catcode `\\12\catcode `\$12\catcode
		`\&12\catcode `\#12\catcode `\^12\catcode `\_12\catcode `\%12\relax}%
	\providecommand \@@startlink[1]{}%
	\providecommand \@@endlink[0]{}%
	\providecommand \url  [0]{\begingroup\@sanitize@url \@url }%
	\providecommand \@url [1]{\endgroup\@href {#1}{\urlprefix }}%
	\providecommand \urlprefix  [0]{URL }%
	\providecommand \Eprint [0]{\href }%
	\providecommand \doibase [0]{http://dx.doi.org/}%
	\providecommand \selectlanguage [0]{\@gobble}%
	\providecommand \bibinfo  [0]{\@secondoftwo}%
	\providecommand \bibfield  [0]{\@secondoftwo}%
	\providecommand \translation [1]{[#1]}%
	\providecommand \BibitemOpen [0]{}%
	\providecommand \bibitemStop [0]{}%
	\providecommand \bibitemNoStop [0]{.\EOS\space}%
	\providecommand \EOS [0]{\spacefactor3000\relax}%
	\providecommand \BibitemShut  [1]{\csname bibitem#1\endcsname}%
	\let\auto@bib@innerbib\@empty
	\bibitem [{\citenamefont {Anderson}(1958)}]{Anderson1958}%
	\BibitemOpen
	\bibfield  {author} {\bibinfo {author} {\bibfnamefont {P.~W.}\ \bibnamefont
			{Anderson}},\ }\bibfield  {title} {\enquote {\bibinfo {title} {Absence of
				diffusion in certain random lattices},}\ }\href {\doibase
		10.1103/PhysRev.109.1492} {\bibfield  {journal} {\bibinfo  {journal} {Phys.
				Rev.}\ }\textbf {\bibinfo {volume} {109}},\ \bibinfo {pages} {1492} (\bibinfo
		{year} {1958})}\BibitemShut {NoStop}%
	\bibitem [{\citenamefont {Anderson}(1985)}]{Anderson1985}%
	\BibitemOpen
	\bibfield  {author} {\bibinfo {author} {\bibfnamefont {P.~W.}\ \bibnamefont
			{Anderson}},\ }\bibfield  {title} {\enquote {\bibinfo {title} {The question
				of classical localization: {A} theory of white paint?}}\ }\href {\doibase
		10.1080/13642818508240619} {\bibfield  {journal} {\bibinfo  {journal}
			{Philos. Mag. B}\ }\textbf {\bibinfo {volume} {52}},\ \bibinfo {pages} {505}
		(\bibinfo {year} {1985})}\BibitemShut {NoStop}%
	\bibitem [{\citenamefont {Lagendijk}\ \emph {et~al.}(2009)\citenamefont
		{Lagendijk}, \citenamefont {{van Tiggelen}},\ and\ \citenamefont
		{Wiersma}}]{Lagendijk2009}%
	\BibitemOpen
	\bibfield  {author} {\bibinfo {author} {\bibfnamefont {A.}~\bibnamefont
			{Lagendijk}}, \bibinfo {author} {\bibfnamefont {B.}~\bibnamefont {{van
					Tiggelen}}}, \ and\ \bibinfo {author} {\bibfnamefont {D.~S.}\ \bibnamefont
			{Wiersma}},\ }\bibfield  {title} {\enquote {\bibinfo {title} {Fifty years of
				{A}nderson localization},}\ }\href {\doibase 10.1063/1.3206091} {\bibfield
		{journal} {\bibinfo  {journal} {Phys. Today}\ }\textbf {\bibinfo {volume}
			{62}},\ \bibinfo {pages} {24} (\bibinfo {year} {2009})}\BibitemShut {NoStop}%
	\bibitem [{\citenamefont {John}(1984)}]{John1984}%
	\BibitemOpen
	\bibfield  {author} {\bibinfo {author} {\bibfnamefont {S.}~\bibnamefont
			{John}},\ }\bibfield  {title} {\enquote {\bibinfo {title} {Electromagnetic
				absorption in a disordered medium near a photon mobility edge},}\ }\href
	{\doibase 10.1103/PhysRevLett.53.2169} {\bibfield  {journal} {\bibinfo
			{journal} {Phys. Rev. Lett.}\ }\textbf {\bibinfo {volume} {53}},\ \bibinfo
		{pages} {2169} (\bibinfo {year} {1984})}\BibitemShut {NoStop}%
	\bibitem [{\citenamefont {John}(1987)}]{John1987}%
	\BibitemOpen
	\bibfield  {author} {\bibinfo {author} {\bibfnamefont {S.}~\bibnamefont
			{John}},\ }\bibfield  {title} {\enquote {\bibinfo {title} {Strong
				localization of photons in certain disordered dielectric superlattices},}\
	}\href {\doibase 10.1103/PhysRevLett.58.2486} {\bibfield  {journal} {\bibinfo
			{journal} {Phys. Rev. Lett.}\ }\textbf {\bibinfo {volume} {58}},\ \bibinfo
		{pages} {2486} (\bibinfo {year} {1987})}\BibitemShut {NoStop}%
	\bibitem [{\citenamefont {{De Raedt}}\ \emph {et~al.}(1989)\citenamefont {{De
				Raedt}}, \citenamefont {Lagendijk},\ and\ \citenamefont {{de
				Vries}}}]{DeRaedt1989}%
	\BibitemOpen
	\bibfield  {author} {\bibinfo {author} {\bibfnamefont {H.}~\bibnamefont {{De
					Raedt}}}, \bibinfo {author} {\bibfnamefont {A.}~\bibnamefont {Lagendijk}}, \
		and\ \bibinfo {author} {\bibfnamefont {P.}~\bibnamefont {{de Vries}}},\
	}\bibfield  {title} {\enquote {\bibinfo {title} {Transverse localization of
				light},}\ }\href {\doibase 10.1103/PhysRevLett.62.47} {\bibfield  {journal}
		{\bibinfo  {journal} {Phys. Rev. Lett.}\ }\textbf {\bibinfo {volume} {62}},\
		\bibinfo {pages} {47} (\bibinfo {year} {1989})}\BibitemShut {NoStop}%
	\bibitem [{\citenamefont {Yannopapas}\ \emph {et~al.}(2003)\citenamefont
		{Yannopapas}, \citenamefont {Modinos},\ and\ \citenamefont
		{Stefanou}}]{Yannopapas2003}%
	\BibitemOpen
	\bibfield  {author} {\bibinfo {author} {\bibfnamefont {V.}~\bibnamefont
			{Yannopapas}}, \bibinfo {author} {\bibfnamefont {A.}~\bibnamefont {Modinos}},
		\ and\ \bibinfo {author} {\bibfnamefont {N.}~\bibnamefont {Stefanou}},\
	}\bibfield  {title} {\enquote {\bibinfo {title} {{A}nderson localization of
				light in inverted opals},}\ }\href {\doibase 10.1103/PhysRevB.68.193205}
	{\bibfield  {journal} {\bibinfo  {journal} {Phys. Rev. B}\ }\textbf {\bibinfo
			{volume} {68}},\ \bibinfo {pages} {193205} (\bibinfo {year}
		{2003})}\BibitemShut {NoStop}%
	\bibitem [{\citenamefont {Schwartz}\ \emph {et~al.}(2007)\citenamefont
		{Schwartz}, \citenamefont {Bartal}, \citenamefont {Fishman},\ and\
		\citenamefont {Segev}}]{Schwartz2007}%
	\BibitemOpen
	\bibfield  {author} {\bibinfo {author} {\bibfnamefont {T.}~\bibnamefont
			{Schwartz}}, \bibinfo {author} {\bibfnamefont {G.}~\bibnamefont {Bartal}},
		\bibinfo {author} {\bibfnamefont {S.}~\bibnamefont {Fishman}}, \ and\
		\bibinfo {author} {\bibfnamefont {M.}~\bibnamefont {Segev}},\ }\bibfield
	{title} {\enquote {\bibinfo {title} {Transport and {A}nderson localization in
				disordered two-dimensional photonic lattices},}\ }\href {\doibase
		10.1038/nature05623} {\bibfield  {journal} {\bibinfo  {journal} {Nature}\
		}\textbf {\bibinfo {volume} {446}},\ \bibinfo {pages} {52} (\bibinfo {year}
		{2007})}\BibitemShut {NoStop}%
	\bibitem [{\citenamefont {Martin}\ \emph {et~al.}(2011)\citenamefont {Martin},
		\citenamefont {{Di Giuseppe}}, \citenamefont {Perez-Leija}, \citenamefont
		{Keil}, \citenamefont {Dreisow}, \citenamefont {Heinrich}, \citenamefont
		{Nolte}, \citenamefont {Szameit}, \citenamefont {Abouraddy}, \citenamefont
		{Christodoulides},\ and\ \citenamefont {Saleh}}]{Martin2011}%
	\BibitemOpen
	\bibfield  {author} {\bibinfo {author} {\bibfnamefont {L.}~\bibnamefont
			{Martin}}, \bibinfo {author} {\bibfnamefont {G.}~\bibnamefont {{Di
					Giuseppe}}}, \bibinfo {author} {\bibfnamefont {A.}~\bibnamefont
			{Perez-Leija}}, \bibinfo {author} {\bibfnamefont {R.}~\bibnamefont {Keil}},
		\bibinfo {author} {\bibfnamefont {F.}~\bibnamefont {Dreisow}}, \bibinfo
		{author} {\bibfnamefont {M.}~\bibnamefont {Heinrich}}, \bibinfo {author}
		{\bibfnamefont {S.}~\bibnamefont {Nolte}}, \bibinfo {author} {\bibfnamefont
			{A.}~\bibnamefont {Szameit}}, \bibinfo {author} {\bibfnamefont {A.~F.}\
			\bibnamefont {Abouraddy}}, \bibinfo {author} {\bibfnamefont {D.~N.}\
			\bibnamefont {Christodoulides}}, \ and\ \bibinfo {author} {\bibfnamefont
			{B.~E.~A.}\ \bibnamefont {Saleh}},\ }\bibfield  {title} {\enquote {\bibinfo
			{title} {{A}nderson localization in optical waveguide arrays with
				off-diagonal coupling disorder},}\ }\href {\doibase 10.1364/OE.19.013636}
	{\bibfield  {journal} {\bibinfo  {journal} {Opt. Express}\ }\textbf {\bibinfo
			{volume} {19}},\ \bibinfo {pages} {13636} (\bibinfo {year}
		{2011})}\BibitemShut {NoStop}%
	\bibitem [{\citenamefont {Mafi}(2015)}]{Mafi2015}%
	\BibitemOpen
	\bibfield  {author} {\bibinfo {author} {\bibfnamefont {A.}~\bibnamefont
			{Mafi}},\ }\bibfield  {title} {\enquote {\bibinfo {title} {Transverse
				{A}nderson localization of light: a tutorial},}\ }\href {\doibase
		10.1364/AOP.7.000459} {\bibfield  {journal} {\bibinfo  {journal} {Adv. Opt.
				Photon.}\ }\textbf {\bibinfo {volume} {7}},\ \bibinfo {pages} {459} (\bibinfo
		{year} {2015})},\ \Eprint {http://arxiv.org/abs/1505.01109} {arXiv:1505.01109
		[physics.optics]} \BibitemShut {NoStop}%
	\bibitem [{\citenamefont {Lee}\ and\ \citenamefont {Fisher}(1981)}]{Lee1981}%
	\BibitemOpen
	\bibfield  {author} {\bibinfo {author} {\bibfnamefont {P.~A.}\ \bibnamefont
			{Lee}}\ and\ \bibinfo {author} {\bibfnamefont {D.~S.}\ \bibnamefont
			{Fisher}},\ }\bibfield  {title} {\enquote {\bibinfo {title} {{A}nderson
				localization in two dimensions},}\ }\href {\doibase
		10.1103/PhysRevLett.47.882} {\bibfield  {journal} {\bibinfo  {journal} {Phys.
				Rev. Lett.}\ }\textbf {\bibinfo {volume} {47}},\ \bibinfo {pages} {882}
		(\bibinfo {year} {1981})}\BibitemShut {NoStop}%
	\bibitem [{\citenamefont {Gornyi}\ \emph {et~al.}(2005)\citenamefont {Gornyi},
		\citenamefont {Mirlin},\ and\ \citenamefont {Polyakov}}]{Gornyi2005}%
	\BibitemOpen
	\bibfield  {author} {\bibinfo {author} {\bibfnamefont {I.~V.}\ \bibnamefont
			{Gornyi}}, \bibinfo {author} {\bibfnamefont {A.~D.}\ \bibnamefont {Mirlin}},
		\ and\ \bibinfo {author} {\bibfnamefont {D.~G.}\ \bibnamefont {Polyakov}},\
	}\bibfield  {title} {\enquote {\bibinfo {title} {Interacting electrons in
				disordered wires: {A}nderson localization and low-{T} transport},}\ }\href
	{\doibase 10.1103/PhysRevLett.95.206603} {\bibfield  {journal} {\bibinfo
			{journal} {Phys. Rev. Lett.}\ }\textbf {\bibinfo {volume} {95}},\ \bibinfo
		{pages} {206603} (\bibinfo {year} {2005})},\ \Eprint
	{http://arxiv.org/abs/cond-mat/0506411} {arXiv:cond-mat/0506411} \BibitemShut
	{NoStop}%
	\bibitem [{\citenamefont {Ying}\ \emph {et~al.}(2016)\citenamefont {Ying},
		\citenamefont {Gu}, \citenamefont {Chen}, \citenamefont {Wang}, \citenamefont
		{Jin}, \citenamefont {Zhao}, \citenamefont {Zhang},\ and\ \citenamefont
		{Chen}}]{Ying2016}%
	\BibitemOpen
	\bibfield  {author} {\bibinfo {author} {\bibfnamefont {T.}~\bibnamefont
			{Ying}}, \bibinfo {author} {\bibfnamefont {Y.}~\bibnamefont {Gu}}, \bibinfo
		{author} {\bibfnamefont {X.}~\bibnamefont {Chen}}, \bibinfo {author}
		{\bibfnamefont {X.}~\bibnamefont {Wang}}, \bibinfo {author} {\bibfnamefont
			{S.}~\bibnamefont {Jin}}, \bibinfo {author} {\bibfnamefont {L.}~\bibnamefont
			{Zhao}}, \bibinfo {author} {\bibfnamefont {W.}~\bibnamefont {Zhang}}, \ and\
		\bibinfo {author} {\bibfnamefont {X.}~\bibnamefont {Chen}},\ }\bibfield
	{title} {\enquote {\bibinfo {title} {{A}nderson localization of electrons in
				single crystals: {Li}$_{x}${Fe}$_{8}${Se}$_{7}$},}\ }\href {\doibase
		10.1126/sciadv.1501283} {\bibfield  {journal} {\bibinfo  {journal} {Sci.
				Adv.}\ }\textbf {\bibinfo {volume} {2}},\ \bibinfo {pages} {e1501283}
		(\bibinfo {year} {2016})}\BibitemShut {NoStop}%
	\bibitem [{\citenamefont {Roati}\ \emph {et~al.}(2008)\citenamefont {Roati},
		\citenamefont {D'Errico}, \citenamefont {Fallani}, \citenamefont {Fattori},
		\citenamefont {Fort}, \citenamefont {Zaccanti}, \citenamefont {Modugno},
		\citenamefont {Modugno},\ and\ \citenamefont {Inguscio}}]{Roati2008}%
	\BibitemOpen
	\bibfield  {author} {\bibinfo {author} {\bibfnamefont {G.}~\bibnamefont
			{Roati}}, \bibinfo {author} {\bibfnamefont {C.}~\bibnamefont {D'Errico}},
		\bibinfo {author} {\bibfnamefont {L.}~\bibnamefont {Fallani}}, \bibinfo
		{author} {\bibfnamefont {M.}~\bibnamefont {Fattori}}, \bibinfo {author}
		{\bibfnamefont {C.}~\bibnamefont {Fort}}, \bibinfo {author} {\bibfnamefont
			{M.}~\bibnamefont {Zaccanti}}, \bibinfo {author} {\bibfnamefont
			{G.}~\bibnamefont {Modugno}}, \bibinfo {author} {\bibfnamefont
			{M.}~\bibnamefont {Modugno}}, \ and\ \bibinfo {author} {\bibfnamefont
			{M.}~\bibnamefont {Inguscio}},\ }\bibfield  {title} {\enquote {\bibinfo
			{title} {{A}nderson localization of a non-interacting {B}ose-{E}instein
				condensate},}\ }\href {\doibase 10.1038/nature07071} {\bibfield  {journal}
		{\bibinfo  {journal} {Nature}\ }\textbf {\bibinfo {volume} {453}},\ \bibinfo
		{pages} {895} (\bibinfo {year} {2008})},\ \Eprint
	{http://arxiv.org/abs/0804.2609} {arXiv:0804.2609 [cond-mat.dis-nn]}
	\BibitemShut {NoStop}%
	\bibitem [{\citenamefont {Plenio}\ and\ \citenamefont
		{Huelga}(2008)}]{Plenio2008}%
	\BibitemOpen
	\bibfield  {author} {\bibinfo {author} {\bibfnamefont {M.~B.}\ \bibnamefont
			{Plenio}}\ and\ \bibinfo {author} {\bibfnamefont {S.~F.}\ \bibnamefont
			{Huelga}},\ }\bibfield  {title} {\enquote {\bibinfo {title}
			{Dephasing-assisted transport: quantum networks and biomolecules},}\ }\href
	{\doibase 10.1063/1.3223548} {\bibfield  {journal} {\bibinfo  {journal} {New
				J. Phys.}\ }\textbf {\bibinfo {volume} {10}},\ \bibinfo {pages} {113019}
		(\bibinfo {year} {2008})},\ \Eprint {http://arxiv.org/abs/0807.4902}
	{arXiv:0807.4902 [quant-ph]} \BibitemShut {NoStop}%
	\bibitem [{\citenamefont {Caruso}\ \emph {et~al.}(2009)\citenamefont {Caruso},
		\citenamefont {Chin}, \citenamefont {Datta}, \citenamefont {Huelga},\ and\
		\citenamefont {Plenio}}]{Caruso2009}%
	\BibitemOpen
	\bibfield  {author} {\bibinfo {author} {\bibfnamefont {F.}~\bibnamefont
			{Caruso}}, \bibinfo {author} {\bibfnamefont {A.~W.}\ \bibnamefont {Chin}},
		\bibinfo {author} {\bibfnamefont {A.}~\bibnamefont {Datta}}, \bibinfo
		{author} {\bibfnamefont {S.~F.}\ \bibnamefont {Huelga}}, \ and\ \bibinfo
		{author} {\bibfnamefont {M.~B.}\ \bibnamefont {Plenio}},\ }\bibfield  {title}
	{\enquote {\bibinfo {title} {Highly efficient energy excitation transfer in
				light-harvesting complexes: the fundamental role of noise-assisted
				transport},}\ }\href {\doibase 10.1063/1.3223548} {\bibfield  {journal}
		{\bibinfo  {journal} {J. Chem. Phys.}\ }\textbf {\bibinfo {volume} {131}},\
		\bibinfo {pages} {105106} (\bibinfo {year} {2009})},\ \Eprint
	{http://arxiv.org/abs/0901.4454} {arXiv:0901.4454 [quant-ph]} \BibitemShut
	{NoStop}%
	\bibitem [{\citenamefont {Viciani}\ \emph {et~al.}(2015)\citenamefont
		{Viciani}, \citenamefont {Lima}, \citenamefont {Bellini},\ and\ \citenamefont
		{Caruso}}]{Viciani2015}%
	\BibitemOpen
	\bibfield  {author} {\bibinfo {author} {\bibfnamefont {S.}~\bibnamefont
			{Viciani}}, \bibinfo {author} {\bibfnamefont {M.}~\bibnamefont {Lima}},
		\bibinfo {author} {\bibfnamefont {M.}~\bibnamefont {Bellini}}, \ and\
		\bibinfo {author} {\bibfnamefont {F.}~\bibnamefont {Caruso}},\ }\bibfield
	{title} {\enquote {\bibinfo {title} {Observation of noise-assisted transport
				in an all-optical cavity-based network},}\ }\href {\doibase
		10.1103/PhysRevLett.115.083601} {\bibfield  {journal} {\bibinfo  {journal}
			{Phys. Rev. Lett.}\ }\textbf {\bibinfo {volume} {115}},\ \bibinfo {pages}
		{083601} (\bibinfo {year} {2015})},\ \Eprint
	{http://arxiv.org/abs/1504.04809} {arXiv:1504.04809 [quant.phys]}
	\BibitemShut {NoStop}%
	\bibitem [{\citenamefont {Viciani}\ \emph {et~al.}(2016)\citenamefont
		{Viciani}, \citenamefont {Gherardini}, \citenamefont {Lima}, \citenamefont
		{Bellini},\ and\ \citenamefont {Caruso}}]{Viciani2016}%
	\BibitemOpen
	\bibfield  {author} {\bibinfo {author} {\bibfnamefont {S.}~\bibnamefont
			{Viciani}}, \bibinfo {author} {\bibfnamefont {S.}~\bibnamefont {Gherardini}},
		\bibinfo {author} {\bibfnamefont {M.}~\bibnamefont {Lima}}, \bibinfo {author}
		{\bibfnamefont {M.}~\bibnamefont {Bellini}}, \ and\ \bibinfo {author}
		{\bibfnamefont {F.}~\bibnamefont {Caruso}},\ }\bibfield  {title} {\enquote
		{\bibinfo {title} {Disorder and dephasing as control knobs for light
				transport in optical fiber cavity networks},}\ }\href {\doibase
		10.1038/srep37791} {\bibfield  {journal} {\bibinfo  {journal} {Sci. Rep.}\
		}\textbf {\bibinfo {volume} {6}},\ \bibinfo {pages} {37791} (\bibinfo {year}
		{2016})},\ \Eprint {http://arxiv.org/abs/1806.02856} {arXiv:1806.02856
		[quant.phys]} \BibitemShut {NoStop}%
	\bibitem [{\citenamefont {Biggerstaff}\ \emph {et~al.}(2016)\citenamefont
		{Biggerstaff}, \citenamefont {Heilmann}, \citenamefont {Zecevik},
		\citenamefont {Gr\"afe}, \citenamefont {Broome}, \citenamefont {Fedrizzi},
		\citenamefont {Nolte}, \citenamefont {Szameit}, \citenamefont {White},\ and\
		\citenamefont {Kassal}}]{Biggerstaff2016}%
	\BibitemOpen
	\bibfield  {author} {\bibinfo {author} {\bibfnamefont {D.~N.}\ \bibnamefont
			{Biggerstaff}}, \bibinfo {author} {\bibfnamefont {R.}~\bibnamefont
			{Heilmann}}, \bibinfo {author} {\bibfnamefont {A.~A.}\ \bibnamefont
			{Zecevik}}, \bibinfo {author} {\bibfnamefont {M.}~\bibnamefont {Gr\"afe}},
		\bibinfo {author} {\bibfnamefont {M.~A.}\ \bibnamefont {Broome}}, \bibinfo
		{author} {\bibfnamefont {A.}~\bibnamefont {Fedrizzi}}, \bibinfo {author}
		{\bibfnamefont {S.}~\bibnamefont {Nolte}}, \bibinfo {author} {\bibfnamefont
			{A.}~\bibnamefont {Szameit}}, \bibinfo {author} {\bibfnamefont {A.~G.}\
			\bibnamefont {White}}, \ and\ \bibinfo {author} {\bibfnamefont
			{I.}~\bibnamefont {Kassal}},\ }\bibfield  {title} {\enquote {\bibinfo {title}
			{Enhancing coherent transport in a photonic network using controllable
				decoherence},}\ }\href {\doibase 10.1038/ncomms11282} {\bibfield  {journal}
		{\bibinfo  {journal} {Nat. Commun.}\ }\textbf {\bibinfo {volume} {7}},\
		\bibinfo {pages} {11282} (\bibinfo {year} {2016})},\ \Eprint
	{http://arxiv.org/abs/1504.06152} {arXiv:1504.06152 [quant.phys]}
	\BibitemShut {NoStop}%
	\bibitem [{\citenamefont {Caruso}\ \emph {et~al.}(2016)\citenamefont {Caruso},
		\citenamefont {Crespi}, \citenamefont {Ciriolo}, \citenamefont {Sciarrino},\
		and\ \citenamefont {Osellame}}]{Caruso2016}%
	\BibitemOpen
	\bibfield  {author} {\bibinfo {author} {\bibfnamefont {F.}~\bibnamefont
			{Caruso}}, \bibinfo {author} {\bibfnamefont {A.}~\bibnamefont {Crespi}},
		\bibinfo {author} {\bibfnamefont {A.~G.}\ \bibnamefont {Ciriolo}}, \bibinfo
		{author} {\bibfnamefont {F.}~\bibnamefont {Sciarrino}}, \ and\ \bibinfo
		{author} {\bibfnamefont {R.}~\bibnamefont {Osellame}},\ }\bibfield  {title}
	{\enquote {\bibinfo {title} {Fast escape of a quantum walker from an
				integrated photonic maze},}\ }\href {\doibase 10.1038/ncomms11682} {\bibfield
		{journal} {\bibinfo  {journal} {Nat. Commun.}\ }\textbf {\bibinfo {volume}
			{7}},\ \bibinfo {pages} {11682} (\bibinfo {year} {2016})},\ \Eprint
	{http://arxiv.org/abs/1501.06438} {arXiv:1501.06438 [quant.phys]}
	\BibitemShut {NoStop}%
	\bibitem [{\citenamefont {Intravaia}\ \emph {et~al.}(2019)\citenamefont
		{Intravaia}, \citenamefont {Dalvit},\ and\ \citenamefont
		{Busch}}]{Intravaia2019}%
	\BibitemOpen
	\bibfield  {author} {\bibinfo {author} {\bibfnamefont {F.}~\bibnamefont
			{Intravaia}}, \bibinfo {author} {\bibfnamefont {D.~A.~R.}\ \bibnamefont
			{Dalvit}}, \ and\ \bibinfo {author} {\bibfnamefont {K.}~\bibnamefont
			{Busch}},\ }\bibfield  {title} {\enquote {\bibinfo {title}
			{Fluctuation-induced phenomena in photonic systems: introduction},}\ }\href
	{\doibase 10.1364/JOSAB.36.00FIP1} {\bibfield  {journal} {\bibinfo  {journal}
			{J. Opt. Soc. Am. B}\ }\textbf {\bibinfo {volume} {36}},\ \bibinfo {pages}
		{FIP1} (\bibinfo {year} {2019})}\BibitemShut {NoStop}%
	\bibitem [{\citenamefont {Takenaga}\ \emph {et~al.}(2011)\citenamefont
		{Takenaga}, \citenamefont {Arakawa}, \citenamefont {Tanigawa}, \citenamefont
		{Guan}, \citenamefont {Matsuo}, \citenamefont {Saitoh},\ and\ \citenamefont
		{Koshiba}}]{Takenaga2011}%
	\BibitemOpen
	\bibfield  {author} {\bibinfo {author} {\bibfnamefont {K.}~\bibnamefont
			{Takenaga}}, \bibinfo {author} {\bibfnamefont {Y.}~\bibnamefont {Arakawa}},
		\bibinfo {author} {\bibfnamefont {S.}~\bibnamefont {Tanigawa}}, \bibinfo
		{author} {\bibfnamefont {N.}~\bibnamefont {Guan}}, \bibinfo {author}
		{\bibfnamefont {S.}~\bibnamefont {Matsuo}}, \bibinfo {author} {\bibfnamefont
			{K.}~\bibnamefont {Saitoh}}, \ and\ \bibinfo {author} {\bibfnamefont
			{M.}~\bibnamefont {Koshiba}},\ }\bibfield  {title} {\enquote {\bibinfo
			{title} {An investigation on crosstalk in multi-core fibers by introducing
				random fluctuation along longitudinal direction},}\ }\href {\doibase
		10.1587/transcom.E94.B.409} {\bibfield  {journal} {\bibinfo  {journal} {IEICE
				Trans. Commun.}\ }\textbf {\bibinfo {volume} {E94.B}},\ \bibinfo {pages}
		{409--416} (\bibinfo {year} {2011})}\BibitemShut {NoStop}%
	\bibitem [{\citenamefont {Matsuo}\ \emph
		{et~al.}(2011{\natexlab{a}})\citenamefont {Matsuo}, \citenamefont {Takenaga},
		\citenamefont {Arakawa}, \citenamefont {Sasaki}, \citenamefont {Tanigawa},
		\citenamefont {Saitoh},\ and\ \citenamefont {Koshiba}}]{Matsuo2011a}%
	\BibitemOpen
	\bibfield  {author} {\bibinfo {author} {\bibfnamefont {S.}~\bibnamefont
			{Matsuo}}, \bibinfo {author} {\bibfnamefont {K.}~\bibnamefont {Takenaga}},
		\bibinfo {author} {\bibfnamefont {Y.}~\bibnamefont {Arakawa}}, \bibinfo
		{author} {\bibfnamefont {Y.}~\bibnamefont {Sasaki}}, \bibinfo {author}
		{\bibfnamefont {S.}~\bibnamefont {Tanigawa}}, \bibinfo {author}
		{\bibfnamefont {K.}~\bibnamefont {Saitoh}}, \ and\ \bibinfo {author}
		{\bibfnamefont {M.}~\bibnamefont {Koshiba}},\ }\bibfield  {title} {\enquote
		{\bibinfo {title} {Crosstalk behavior of cores in multi-core fiber under bent
				condition},}\ }\href {\doibase 10.1587/elex.8.385} {\bibfield  {journal}
		{\bibinfo  {journal} {IEICE Electron. Expr.}\ }\textbf {\bibinfo {volume}
			{8}},\ \bibinfo {pages} {385--390} (\bibinfo {year}
		{2011}{\natexlab{a}})}\BibitemShut {NoStop}%
	\bibitem [{\citenamefont {Matsuo}\ \emph
		{et~al.}(2011{\natexlab{b}})\citenamefont {Matsuo}, \citenamefont {Takenaga},
		\citenamefont {Arakawa}, \citenamefont {Sasaki}, \citenamefont {Tanigawa},
		\citenamefont {Saitoh},\ and\ \citenamefont {Koshiba}}]{Matsuo2011b}%
	\BibitemOpen
	\bibfield  {author} {\bibinfo {author} {\bibfnamefont {S.}~\bibnamefont
			{Matsuo}}, \bibinfo {author} {\bibfnamefont {K.}~\bibnamefont {Takenaga}},
		\bibinfo {author} {\bibfnamefont {Y.}~\bibnamefont {Arakawa}}, \bibinfo
		{author} {\bibfnamefont {Y.}~\bibnamefont {Sasaki}}, \bibinfo {author}
		{\bibfnamefont {S.}~\bibnamefont {Tanigawa}}, \bibinfo {author}
		{\bibfnamefont {K.}~\bibnamefont {Saitoh}}, \ and\ \bibinfo {author}
		{\bibfnamefont {M.}~\bibnamefont {Koshiba}},\ }\bibfield  {title} {\enquote
		{\bibinfo {title} {Crosstalk behavior of multi-core fiber with structural
				parameter drift in longitudinal direction},}\ }\href {\doibase
		10.1587/elex.8.1419} {\bibfield  {journal} {\bibinfo  {journal} {IEICE
				Electron. Expr.}\ }\textbf {\bibinfo {volume} {8}},\ \bibinfo {pages}
		{1419--1424} (\bibinfo {year} {2011}{\natexlab{b}})}\BibitemShut {NoStop}%
	\bibitem [{\citenamefont {Fini}\ \emph {et~al.}(2012)\citenamefont {Fini},
		\citenamefont {Zhu}, \citenamefont {Taunay}, \citenamefont {Yan},\ and\
		\citenamefont {Abedin}}]{Fini2012}%
	\BibitemOpen
	\bibfield  {author} {\bibinfo {author} {\bibfnamefont {J.~M.}\ \bibnamefont
			{Fini}}, \bibinfo {author} {\bibfnamefont {B.}~\bibnamefont {Zhu}}, \bibinfo
		{author} {\bibfnamefont {T.~F.}\ \bibnamefont {Taunay}}, \bibinfo {author}
		{\bibfnamefont {M.~F.}\ \bibnamefont {Yan}}, \ and\ \bibinfo {author}
		{\bibfnamefont {K.~S.}\ \bibnamefont {Abedin}},\ }\bibfield  {title}
	{\enquote {\bibinfo {title} {Crosstalk in multicore fibers with randomness:
				gradual drift vs. short-length variations},}\ }\href {\doibase
		10.1364/OE.20.000949} {\bibfield  {journal} {\bibinfo  {journal} {Opt.
				Express}\ }\textbf {\bibinfo {volume} {20}},\ \bibinfo {pages} {949}
		(\bibinfo {year} {2012})}\BibitemShut {NoStop}%
	\bibitem [{\citenamefont {Fini}\ \emph {et~al.}(2010)\citenamefont {Fini},
		\citenamefont {Zhu}, \citenamefont {Taunay},\ and\ \citenamefont
		{Yan}}]{Fini2010}%
	\BibitemOpen
	\bibfield  {author} {\bibinfo {author} {\bibfnamefont {John~M.}\ \bibnamefont
			{Fini}}, \bibinfo {author} {\bibfnamefont {Benyuan}\ \bibnamefont {Zhu}},
		\bibinfo {author} {\bibfnamefont {Thierry~F.}\ \bibnamefont {Taunay}}, \ and\
		\bibinfo {author} {\bibfnamefont {Man~F.}\ \bibnamefont {Yan}},\ }\bibfield
	{title} {\enquote {\bibinfo {title} {Statistics of crosstalk in bent
				multicore fibers},}\ }\href {\doibase 10.1364/OE.18.015122} {\bibfield
		{journal} {\bibinfo  {journal} {Opt. Express}\ }\textbf {\bibinfo {volume}
			{18}},\ \bibinfo {pages} {15122} (\bibinfo {year} {2010})}\BibitemShut
	{NoStop}%
	\bibitem [{\citenamefont {Li}\ \emph {et~al.}(2015)\citenamefont {Li},
		\citenamefont {Li},\ and\ \citenamefont {Modavis}}]{Li2015}%
	\BibitemOpen
	\bibfield  {author} {\bibinfo {author} {\bibfnamefont {M.-J.}\ \bibnamefont
			{Li}}, \bibinfo {author} {\bibfnamefont {S.}~\bibnamefont {Li}}, \ and\
		\bibinfo {author} {\bibfnamefont {R.~A.}\ \bibnamefont {Modavis}},\
	}\bibfield  {title} {\enquote {\bibinfo {title} {Coupled mode analysis of
				crosstalk in multicore fiber with random perturbations},}\ }in\ \href
	{\doibase 10.1364/OFC.2015.W2A.35} {\emph {\bibinfo {booktitle} {Optical
				Fiber Communication Conference}}}\ (\bibinfo  {publisher} {Optical Society of
		America},\ \bibinfo {year} {2015})\ p.\ \bibinfo {pages} {W2A.35}\BibitemShut
	{NoStop}%
	\bibitem [{\citenamefont {Chen}\ and\ \citenamefont
		{Keutzer}(1999)}]{Chen1999}%
	\BibitemOpen
	\bibfield  {author} {\bibinfo {author} {\bibfnamefont {P.}~\bibnamefont
			{Chen}}\ and\ \bibinfo {author} {\bibfnamefont {K.}~\bibnamefont {Keutzer}},\
	}\bibfield  {title} {\enquote {\bibinfo {title} {Towards true crosstalk noise
				analysis},}\ }in\ \href@noop {} {\emph {\bibinfo {booktitle} {Proceedings of
				the 1999 {IEEE/ACM} International Conference on Computer-Aided Design}}},\
	\bibinfo {series and number} {ICCAD '99}\ (\bibinfo  {publisher} {IEEE
		Press},\ \bibinfo {year} {1999})\ p.\ \bibinfo {pages} {132}\BibitemShut
	{NoStop}%
	\bibitem [{\citenamefont {{Jaramillo \'Avila}}\ \emph
		{et~al.}(2019{\natexlab{a}})\citenamefont {{Jaramillo \'Avila}},
		\citenamefont {{Naya Hern\'andez}}, \citenamefont {{Toxqui Rodr\'iguez}},\
		and\ \citenamefont {Rodr\'iguez-Lara}}]{Jaramillo2019a}%
	\BibitemOpen
	\bibfield  {author} {\bibinfo {author} {\bibfnamefont {B.}~\bibnamefont
			{{Jaramillo \'Avila}}}, \bibinfo {author} {\bibfnamefont {J.}~\bibnamefont
			{{Naya Hern\'andez}}}, \bibinfo {author} {\bibfnamefont {S.~Ma.}\
			\bibnamefont {{Toxqui Rodr\'iguez}}}, \ and\ \bibinfo {author} {\bibfnamefont
			{B.~M.}\ \bibnamefont {Rodr\'iguez-Lara}},\ }\bibfield  {title} {\enquote
		{\bibinfo {title} {Optimal crosstalk suppression in multicore fibers},}\
	}\href {\doibase 10.1038/s41598-019-51854-x} {\bibfield  {journal} {\bibinfo
			{journal} {Sci. Rep.}\ }\textbf {\bibinfo {volume} {9}},\ \bibinfo {pages}
		{15737} (\bibinfo {year} {2019}{\natexlab{a}})},\ \Eprint
	{http://arxiv.org/abs/1905.09416} {arXiv:1905.09416 [physics.optics]}
	\BibitemShut {NoStop}%
	\bibitem [{\citenamefont {Snyder}(1972)}]{Snyder1972}%
	\BibitemOpen
	\bibfield  {author} {\bibinfo {author} {\bibfnamefont {A.~W.}\ \bibnamefont
			{Snyder}},\ }\bibfield  {title} {\enquote {\bibinfo {title} {Coupled-mode
				theory for optical fibers},}\ }\href {\doibase 10.1364/JOSA.62.001267}
	{\bibfield  {journal} {\bibinfo  {journal} {J. Opt. Soc. Am.}\ }\textbf
		{\bibinfo {volume} {62}},\ \bibinfo {pages} {1267} (\bibinfo {year}
		{1972})}\BibitemShut {NoStop}%
	\bibitem [{\citenamefont {McIntyre}\ and\ \citenamefont
		{Snyder}(1973)}]{McIntyre1973}%
	\BibitemOpen
	\bibfield  {author} {\bibinfo {author} {\bibfnamefont {P.~D.}\ \bibnamefont
			{McIntyre}}\ and\ \bibinfo {author} {\bibfnamefont {A.~W.}\ \bibnamefont
			{Snyder}},\ }\bibfield  {title} {\enquote {\bibinfo {title} {Power transfer
				between optical fibers},}\ }\href {\doibase 10.1364/josa.63.001518}
	{\bibfield  {journal} {\bibinfo  {journal} {J. Opt. Soc. Am.}\ }\textbf
		{\bibinfo {volume} {63}},\ \bibinfo {pages} {1518} (\bibinfo {year}
		{1973})}\BibitemShut {NoStop}%
	\bibitem [{\citenamefont {Huang}(1994)}]{Huang1994}%
	\BibitemOpen
	\bibfield  {author} {\bibinfo {author} {\bibfnamefont {W.-P.}\ \bibnamefont
			{Huang}},\ }\bibfield  {title} {\enquote {\bibinfo {title} {Coupled-mode
				theory for optical waveguides: an overview},}\ }\href {\doibase
		10.1364/JOSAA.11.000963} {\bibfield  {journal} {\bibinfo  {journal} {J. Opt.
				Soc. Am. A}\ }\textbf {\bibinfo {volume} {11}},\ \bibinfo {pages} {963}
		(\bibinfo {year} {1994})}\BibitemShut {NoStop}%
	\bibitem [{\citenamefont {Preu}\ \emph {et~al.}(2008)\citenamefont {Preu},
		\citenamefont {Schwefel}, \citenamefont {Malzer}, \citenamefont {D\"ohler},
		\citenamefont {Wang}, \citenamefont {Hanson}, \citenamefont {Zimmerman},\
		and\ \citenamefont {Gossard}}]{Preu2008}%
	\BibitemOpen
	\bibfield  {author} {\bibinfo {author} {\bibfnamefont {S.}~\bibnamefont
			{Preu}}, \bibinfo {author} {\bibfnamefont {H.~G.~L.}\ \bibnamefont
			{Schwefel}}, \bibinfo {author} {\bibfnamefont {S.}~\bibnamefont {Malzer}},
		\bibinfo {author} {\bibfnamefont {G.~H.}\ \bibnamefont {D\"ohler}}, \bibinfo
		{author} {\bibfnamefont {L.~J.}\ \bibnamefont {Wang}}, \bibinfo {author}
		{\bibfnamefont {M.}~\bibnamefont {Hanson}}, \bibinfo {author} {\bibfnamefont
			{J.~D.}\ \bibnamefont {Zimmerman}}, \ and\ \bibinfo {author} {\bibfnamefont
			{A.~C.}\ \bibnamefont {Gossard}},\ }\bibfield  {title} {\enquote {\bibinfo
			{title} {Coupled whispering gallery mode resonators in the {T}erahertz
				frequency range},}\ }\href {\doibase 10.1364/OE.16.007336} {\bibfield
		{journal} {\bibinfo  {journal} {Opt. Express}\ }\textbf {\bibinfo {volume}
			{16}},\ \bibinfo {pages} {7336} (\bibinfo {year} {2008})}\BibitemShut
	{NoStop}%
	\bibitem [{\citenamefont {Haus}\ and\ \citenamefont {Huang}(1991)}]{Haus1991}%
	\BibitemOpen
	\bibfield  {author} {\bibinfo {author} {\bibfnamefont {H.~A.}\ \bibnamefont
			{Haus}}\ and\ \bibinfo {author} {\bibfnamefont {W.}~\bibnamefont {Huang}},\
	}\bibfield  {title} {\enquote {\bibinfo {title} {Coupled-mode theory},}\
	}\href {\doibase 10.1109/5.104225} {\bibfield  {journal} {\bibinfo  {journal}
			{Proc. IEEE}\ }\textbf {\bibinfo {volume} {79}},\ \bibinfo {pages} {1505}
		(\bibinfo {year} {1991})}\BibitemShut {NoStop}%
	\bibitem [{\citenamefont {Elnaggar}\ \emph {et~al.}(2014)\citenamefont
		{Elnaggar}, \citenamefont {Tervo},\ and\ \citenamefont
		{Mattar}}]{Elnaggar2014}%
	\BibitemOpen
	\bibfield  {author} {\bibinfo {author} {\bibfnamefont {S.~Y.}\ \bibnamefont
			{Elnaggar}}, \bibinfo {author} {\bibfnamefont {R.}~\bibnamefont {Tervo}}, \
		and\ \bibinfo {author} {\bibfnamefont {S.~M.}\ \bibnamefont {Mattar}},\
	}\bibfield  {title} {\enquote {\bibinfo {title} {Coupled modes, frequencies
				and fields of a dielectric resonator and a cavity using coupled mode
				theory},}\ }\href {\doibase 10.1016/j.jmr.2013.10.016} {\bibfield  {journal}
		{\bibinfo  {journal} {J. Magn. Reson.}\ }\textbf {\bibinfo {volume} {238}},\
		\bibinfo {pages} {1} (\bibinfo {year} {2014})}\BibitemShut {NoStop}%
	\bibitem [{\citenamefont {Elnaggar}\ \emph {et~al.}(2015)\citenamefont
		{Elnaggar}, \citenamefont {Tervo},\ and\ \citenamefont
		{Mattar}}]{Elnaggar2015}%
	\BibitemOpen
	\bibfield  {author} {\bibinfo {author} {\bibfnamefont {S.~Y.}\ \bibnamefont
			{Elnaggar}}, \bibinfo {author} {\bibfnamefont {R.~J.}\ \bibnamefont {Tervo}},
		\ and\ \bibinfo {author} {\bibfnamefont {S.~M.}\ \bibnamefont {Mattar}},\
	}\bibfield  {title} {\enquote {\bibinfo {title} {Energy coupled mode theory
				for electromagnetic resonators},}\ }\href {\doibase
		10.1109/TMTT.2015.2434377} {\bibfield  {journal} {\bibinfo  {journal} {IEEE
				T. Microw. Theory}\ }\textbf {\bibinfo {volume} {63}},\ \bibinfo {pages}
		{2115} (\bibinfo {year} {2015})},\ \Eprint {http://arxiv.org/abs/1305.6085}
	{arXiv:1305.6085 [cond-mat.other]} \BibitemShut {NoStop}%
	\bibitem [{\citenamefont {Kim}\ and\ \citenamefont {Ling}(2007)}]{Kim2007}%
	\BibitemOpen
	\bibfield  {author} {\bibinfo {author} {\bibfnamefont {Y.}~\bibnamefont
			{Kim}}\ and\ \bibinfo {author} {\bibfnamefont {H.}~\bibnamefont {Ling}},\
	}\bibfield  {title} {\enquote {\bibinfo {title} {Investigation of coupled
				mode behaviour of electrically small meander antennas},}\ }\href {\doibase
		10.1049/el20072165} {\bibfield  {journal} {\bibinfo  {journal} {Electron.
				Lett.}\ }\textbf {\bibinfo {volume} {43}},\ \bibinfo {pages} {1250} (\bibinfo
		{year} {2007})}\BibitemShut {NoStop}%
	\bibitem [{\citenamefont {Agarwal}\ \emph {et~al.}(2006)\citenamefont
		{Agarwal}, \citenamefont {Sylvester},\ and\ \citenamefont
		{Blaauw}}]{Agarwal2006}%
	\BibitemOpen
	\bibfield  {author} {\bibinfo {author} {\bibfnamefont {K.}~\bibnamefont
			{Agarwal}}, \bibinfo {author} {\bibfnamefont {D.}~\bibnamefont {Sylvester}},
		\ and\ \bibinfo {author} {\bibfnamefont {D.}~\bibnamefont {Blaauw}},\
	}\bibfield  {title} {\enquote {\bibinfo {title} {Modeling and analysis of
				crosstalk noise in coupled {RLC} interconnects},}\ }\href {\doibase
		10.1109/TCAD.2005.855961} {\bibfield  {journal} {\bibinfo  {journal} {IEEE T.
				Comput. Aid. D.}\ }\textbf {\bibinfo {volume} {25}},\ \bibinfo {pages} {892}
		(\bibinfo {year} {2006})}\BibitemShut {NoStop}%
	\bibitem [{\citenamefont {Liu}\ \emph {et~al.}(2005)\citenamefont {Liu},
		\citenamefont {Chang},\ and\ \citenamefont {Craig}}]{Liu2005}%
	\BibitemOpen
	\bibfield  {author} {\bibinfo {author} {\bibfnamefont {Y.}~\bibnamefont
			{Liu}}, \bibinfo {author} {\bibfnamefont {T.}~\bibnamefont {Chang}}, \ and\
		\bibinfo {author} {\bibfnamefont {A.~E.}\ \bibnamefont {Craig}},\ }\bibfield
	{title} {\enquote {\bibinfo {title} {Coupled mode theory for modeling
				microring resonators},}\ }\href {\doibase 10.1117/1.2012503} {\bibfield
		{journal} {\bibinfo  {journal} {Opt. Eng.}\ }\textbf {\bibinfo {volume}
			{44}},\ \bibinfo {pages} {1} (\bibinfo {year} {2005})}\BibitemShut {NoStop}%
	\bibitem [{\citenamefont {Longhi}(2009)}]{Longhi2009}%
	\BibitemOpen
	\bibfield  {author} {\bibinfo {author} {\bibfnamefont {S.}~\bibnamefont
			{Longhi}},\ }\bibfield  {title} {\enquote {\bibinfo {title} {Quantum-optical
				analogies using photonic structures},}\ }\href {\doibase
		10.1002/lpor.200810055} {\bibfield  {journal} {\bibinfo  {journal} {Laser
				Photonics Rev.}\ }\textbf {\bibinfo {volume} {3}},\ \bibinfo {pages} {243}
		(\bibinfo {year} {2009})}\BibitemShut {NoStop}%
	\bibitem [{\citenamefont {Davis}\ \emph {et~al.}(1996)\citenamefont {Davis},
		\citenamefont {Miura}, \citenamefont {Sugimoto},\ and\ \citenamefont
		{Hirao}}]{Davis1996}%
	\BibitemOpen
	\bibfield  {author} {\bibinfo {author} {\bibfnamefont {K.~M.}\ \bibnamefont
			{Davis}}, \bibinfo {author} {\bibfnamefont {K.}~\bibnamefont {Miura}},
		\bibinfo {author} {\bibfnamefont {N.}~\bibnamefont {Sugimoto}}, \ and\
		\bibinfo {author} {\bibfnamefont {K.}~\bibnamefont {Hirao}},\ }\bibfield
	{title} {\enquote {\bibinfo {title} {Writing waveguides in glass with a
				femtosecond laser},}\ }\href {\doibase 10.1364/OL.21.001729} {\bibfield
		{journal} {\bibinfo  {journal} {Opt. Lett.}\ }\textbf {\bibinfo {volume}
			{21}},\ \bibinfo {pages} {1729--1731} (\bibinfo {year} {1996})}\BibitemShut
	{NoStop}%
	\bibitem [{\citenamefont {Bl\"omer}\ \emph {et~al.}(2006)\citenamefont
		{Bl\"omer}, \citenamefont {Szameit}, \citenamefont {Dreisow}, \citenamefont
		{Schreiber}, \citenamefont {Nolte},\ and\ \citenamefont
		{T\"unnermann}}]{Blomer2006}%
	\BibitemOpen
	\bibfield  {author} {\bibinfo {author} {\bibfnamefont {D.}~\bibnamefont
			{Bl\"omer}}, \bibinfo {author} {\bibfnamefont {A.}~\bibnamefont {Szameit}},
		\bibinfo {author} {\bibfnamefont {F.}~\bibnamefont {Dreisow}}, \bibinfo
		{author} {\bibfnamefont {T.}~\bibnamefont {Schreiber}}, \bibinfo {author}
		{\bibfnamefont {S.}~\bibnamefont {Nolte}}, \ and\ \bibinfo {author}
		{\bibfnamefont {A.}~\bibnamefont {T\"unnermann}},\ }\bibfield  {title}
	{\enquote {\bibinfo {title} {Nonlinear refractive index of fs-laser-written
				waveguides in fused silica},}\ }\href {\doibase 10.1364/OE.14.002151}
	{\bibfield  {journal} {\bibinfo  {journal} {Opt. Express}\ }\textbf {\bibinfo
			{volume} {14}},\ \bibinfo {pages} {2151--2157} (\bibinfo {year}
		{2006})}\BibitemShut {NoStop}%
	\bibitem [{\citenamefont {Eaton}\ \emph {et~al.}(2011)\citenamefont {Eaton},
		\citenamefont {Ng}, \citenamefont {Osellame},\ and\ \citenamefont
		{Herman}}]{Eaton2011}%
	\BibitemOpen
	\bibfield  {author} {\bibinfo {author} {\bibfnamefont {S.~M.}\ \bibnamefont
			{Eaton}}, \bibinfo {author} {\bibfnamefont {M.~L.}\ \bibnamefont {Ng}},
		\bibinfo {author} {\bibfnamefont {R.}~\bibnamefont {Osellame}}, \ and\
		\bibinfo {author} {\bibfnamefont {P.~R.}\ \bibnamefont {Herman}},\ }\bibfield
	{title} {\enquote {\bibinfo {title} {High refractive index contrast in fused
				silica waveguides by tightly focused, high-repetition rate femtosecond
				laser},}\ }\href {\doibase 10.1016/j.jnoncrysol.2010.11.082} {\bibfield
		{journal} {\bibinfo  {journal} {J. Non-Cryst. Solids}\ }\textbf {\bibinfo
			{volume} {357}},\ \bibinfo {pages} {2387--2391} (\bibinfo {year}
		{2011})}\BibitemShut {NoStop}%
	\bibitem [{\citenamefont {de~J.~Le\'on-Montiel}\ \emph
		{et~al.}(2015)\citenamefont {de~J.~Le\'on-Montiel}, \citenamefont
		{Quiroz-Ju\'arez}, \citenamefont {Quintero-Torres}, \citenamefont
		{Dom\'inguez-Ju\'arez}, \citenamefont {Moya-Cessa}, \citenamefont {Torres},\
		and\ \citenamefont {Arag\'on}}]{Leon2015}%
	\BibitemOpen
	\bibfield  {author} {\bibinfo {author} {\bibfnamefont {R.}~\bibnamefont
			{de~J.~Le\'on-Montiel}}, \bibinfo {author} {\bibfnamefont {M.~A.}\
			\bibnamefont {Quiroz-Ju\'arez}}, \bibinfo {author} {\bibfnamefont
			{R.}~\bibnamefont {Quintero-Torres}}, \bibinfo {author} {\bibfnamefont
			{J.~L.}\ \bibnamefont {Dom\'inguez-Ju\'arez}}, \bibinfo {author}
		{\bibfnamefont {H.~M.}\ \bibnamefont {Moya-Cessa}}, \bibinfo {author}
		{\bibfnamefont {J.~P.}\ \bibnamefont {Torres}}, \ and\ \bibinfo {author}
		{\bibfnamefont {J.~L.}\ \bibnamefont {Arag\'on}},\ }\bibfield  {title}
	{\enquote {\bibinfo {title} {Noise-assisted energy transport in electrical
				oscillator networks with off-diagonal dynamical disorder},}\ }\href {\doibase
		10.1038/srep17339} {\bibfield  {journal} {\bibinfo  {journal} {Sci. Rep.}\
		}\textbf {\bibinfo {volume} {5}},\ \bibinfo {pages} {17339} (\bibinfo {year}
		{2015})},\ \Eprint {http://arxiv.org/abs/1509.01272} {arXiv:1509.01272
		[physics.class-ph]} \BibitemShut {NoStop}%
	\bibitem [{\citenamefont {Quiroz-Ju\'arez}\ \emph {et~al.}(2016)\citenamefont
		{Quiroz-Ju\'arez}, \citenamefont {Arag\'on}, \citenamefont
		{de~J.~Le\'on-Montiel}, \citenamefont {V\'azquez-Medina}, \citenamefont
		{Dom\'inguez-Ju\'arez},\ and\ \citenamefont {Quintero-Torres}}]{Quiroz2016}%
	\BibitemOpen
	\bibfield  {author} {\bibinfo {author} {\bibfnamefont {M.~A.}\ \bibnamefont
			{Quiroz-Ju\'arez}}, \bibinfo {author} {\bibfnamefont {J.~L.}\ \bibnamefont
			{Arag\'on}}, \bibinfo {author} {\bibfnamefont {R.}~\bibnamefont
			{de~J.~Le\'on-Montiel}}, \bibinfo {author} {\bibfnamefont {R.}~\bibnamefont
			{V\'azquez-Medina}}, \bibinfo {author} {\bibfnamefont {J.~L.}\ \bibnamefont
			{Dom\'inguez-Ju\'arez}}, \ and\ \bibinfo {author} {\bibfnamefont
			{R.}~\bibnamefont {Quintero-Torres}},\ }\bibfield  {title} {\enquote
		{\bibinfo {title} {Emergence of a negative resistance in noisy coupled linear
				oscillators},}\ }\href {\doibase 10.1209/0295-5075/116/50004} {\bibfield
		{journal} {\bibinfo  {journal} {EPL Europhys. Lett.}\ }\textbf {\bibinfo
			{volume} {116}},\ \bibinfo {pages} {50004} (\bibinfo {year}
		{2016})}\BibitemShut {NoStop}%
	\bibitem [{\citenamefont {Quiroz-Ju\'arez}\ \emph {et~al.}(2018)\citenamefont
		{Quiroz-Ju\'arez}, \citenamefont {Arag\'on}, \citenamefont
		{de~J.~Le\'on-Montiel}, \citenamefont {V\'azquez-Medina}, \citenamefont
		{Dom\'inguez-Ju\'arez},\ and\ \citenamefont {Quintero-Torres}}]{Leon2018}%
	\BibitemOpen
	\bibfield  {author} {\bibinfo {author} {\bibfnamefont {M.~A.}\ \bibnamefont
			{Quiroz-Ju\'arez}}, \bibinfo {author} {\bibfnamefont {J.~L.}\ \bibnamefont
			{Arag\'on}}, \bibinfo {author} {\bibfnamefont {R.}~\bibnamefont
			{de~J.~Le\'on-Montiel}}, \bibinfo {author} {\bibfnamefont {R.}~\bibnamefont
			{V\'azquez-Medina}}, \bibinfo {author} {\bibfnamefont {J.~L.}\ \bibnamefont
			{Dom\'inguez-Ju\'arez}}, \ and\ \bibinfo {author} {\bibfnamefont
			{R.}~\bibnamefont {Quintero-Torres}},\ }\bibfield  {title} {\enquote
		{\bibinfo {title} {Observation of slowly decaying eigenmodes without
				exceptional points in {F}loquet dissipative synthetic circuits},}\ }\href
	{\doibase 10.1038/s42005-018-0087-3} {\bibfield  {journal} {\bibinfo
			{journal} {Commun. Phys.}\ }\textbf {\bibinfo {volume} {1}},\ \bibinfo
		{pages} {88} (\bibinfo {year} {2018})},\ \Eprint
	{http://arxiv.org/abs/1805.08393} {arXiv:1805.08393 [physics.class-ph]}
	\BibitemShut {NoStop}%
	\bibitem [{\citenamefont {{Jaramillo \'Avila}}\ \emph
		{et~al.}(2019{\natexlab{b}})\citenamefont {{Jaramillo \'Avila}},
		\citenamefont {{Naya Hern\'andez}}, \citenamefont {{Toxqui Rodr\'iguez}},\
		and\ \citenamefont {Rodr\'iguez-Lara}}]{Jaramillo2019b}%
	\BibitemOpen
	\bibfield  {author} {\bibinfo {author} {\bibfnamefont {B.}~\bibnamefont
			{{Jaramillo \'Avila}}}, \bibinfo {author} {\bibfnamefont {J.}~\bibnamefont
			{{Naya Hern\'andez}}}, \bibinfo {author} {\bibfnamefont {S.~Ma.}\
			\bibnamefont {{Toxqui Rodr\'iguez}}}, \ and\ \bibinfo {author} {\bibfnamefont
			{B.~M.}\ \bibnamefont {Rodr\'iguez-Lara}},\ }\bibfield  {title} {\enquote
		{\bibinfo {title} {Symmetric supermodes in cyclic multicore fibers},}\ }\href
	{\doibase 10.1364/OSAC.2.000515} {\bibfield  {journal} {\bibinfo  {journal}
			{OSA Continuum}\ }\textbf {\bibinfo {volume} {2}},\ \bibinfo {pages}
		{515--522} (\bibinfo {year} {2019}{\natexlab{b}})},\ \Eprint
	{http://arxiv.org/abs/1810.09608} {arXiv:1810.09608 [physics.optics]}
	\BibitemShut {NoStop}%
	\bibitem [{\citenamefont {Shankar}(1994)}]{BookShankarPrinciples}%
	\BibitemOpen
	\bibfield  {author} {\bibinfo {author} {\bibfnamefont {R.}~\bibnamefont
			{Shankar}},\ }\href@noop {} {\emph {\bibinfo {title} {Principles of Quantum
				Mechanics}}}\ (\bibinfo  {publisher} {Plenum Press},\ \bibinfo {year}
	{1994})\BibitemShut {NoStop}%
	\bibitem [{\citenamefont {Thouless}(1974)}]{Thouless1974}%
	\BibitemOpen
	\bibfield  {author} {\bibinfo {author} {\bibfnamefont {D.~J.}\ \bibnamefont
			{Thouless}},\ }\bibfield  {title} {\enquote {\bibinfo {title} {Electrons in
				disordered systems and the theory of localization},}\ }\href {\doibase
		10.1016/0370-1573(74)90029-5} {\bibfield  {journal} {\bibinfo  {journal}
			{Phys. Rep.}\ }\textbf {\bibinfo {volume} {13}},\ \bibinfo {pages} {93}
		(\bibinfo {year} {1974})}\BibitemShut {NoStop}%
	\bibitem [{\citenamefont {Wegner}(1980)}]{Wegner1980}%
	\BibitemOpen
	\bibfield  {author} {\bibinfo {author} {\bibfnamefont {F.}~\bibnamefont
			{Wegner}},\ }\bibfield  {title} {\enquote {\bibinfo {title} {Inverse
				participation ratio in 2+$\epsilon$ dimensions},}\ }\href {\doibase
		10.1007/BF01325284} {\bibfield  {journal} {\bibinfo  {journal} {Z. Phys. B
				Con. Mat.}\ }\textbf {\bibinfo {volume} {36}},\ \bibinfo {pages} {209}
		(\bibinfo {year} {1980})}\BibitemShut {NoStop}%
\end{thebibliography}
%

\end{document}